\documentclass[%
 amsmath,amssymb,
 aps,prd,showpacs,
]{revtex4-1}

\usepackage{bm}
\usepackage[plainpages=false]{hyperref}
\usepackage{graphicx}
\usepackage{multirow}

\usepackage{color}
\usepackage[normalem]{ulem}

\def\la{\mathrel{\mathpalette\fun <}}

\def\fun#1#2{\lower3.6pt\vbox{\baselineskip0pt\lineskip.9pt
  \ialign{$\mathsurround=0pt#1\hfil##\hfil$\crcr#2\crcr\sim\crcr}}}
\def\bepsilon{{\hbox{\boldmath$\epsilon$}}}

\begin{document}

\title{Astrophysical Haloscopes}

\author{G\"unter Sigl}

\affiliation{Universit\"at Hamburg, {II}. Institute for Theoretical Physics, Luruper Chaussee 149, 22761 Hamburg, Germany}

\email[]{guenter.sigl@desy.de}

\begin{abstract}
We compute the fluxes of radio photons from conversion of axion-like particle dark matter in cosmic magnetic fields.
We find that for axion-like particle masses around $10^{-6}\,$eV and effective coupling constants to photons
$g_{a\gamma}\gtrsim10^{-13}\,{\rm GeV}^{-1}$ strongly magnetized nearby stellar winds may give detectable line-like radio
photon signals, although predicted fluxes are highly uncertain due to the poorly known structure of
the magnetic fields. Nevertheless, it may be worth while to conduct a dedicated search in the direction of such sources.
When combined with a possible future laboratory detection of axion-like dark matter such observations may in turn provide
information on the small scale magnetic field structure in such objects.
\end{abstract}


\maketitle

\section{Introduction}
Axion-like particles (ALPs) have developed into an interesting alternative to the WIMP paradigm of cold
dark matter. Originally axions were motivated by the strong CP problem which can be solved by promoting
the CP-violating phase $\theta$, experimentally constrained to be smaller than $\sim10^{-10}$, to a pseudo-scalar
field $a$ via $\theta\to a/f_a$ with $f_a$ an energy scale known as the Peccei-Quinn scale.
The field $a$ is then dynamically driven to zero in a suitable potential which would explain why
the phase $\theta$ essentially vanishes. In such models the axion field couples to the
gluon field strength tensor $G^\alpha_{\mu\nu}$ via a term of the form~\cite{Peccei:1977hh}
\begin{equation}\label{eq:L_aG}
  {\cal L}_{aG}=\frac{\alpha_{\rm s}}{8\pi f_a}\,a\,G^\alpha_{\mu\nu}\tilde G^{\mu\nu}_\alpha\,,
\end{equation}
where $\tilde G^{\mu\nu}_\alpha$ is the dual to $G^\alpha_{\mu\nu}$
and $\alpha_{\rm s}$ is the strong fine structure constant. In QCD axion models the axion mass $m_a$ is related to the
Peccei Quinn scale by~\cite{Weinberg:1977ma,Wilczek:1977pj}
\begin{equation}\label{eq:m_a}
m_a\simeq6\times10^{-6}\,\left(\frac{10^{12}\,{\rm GeV}}{f_a}\right)\,{\rm eV}\,.
\end{equation}

In generalizations of such scenarios to ALPs $f_a$ and $m_a$ are considered as independent parameters and there is
a coupling term to photons of similar shape to the ALP-gluon coupling term. Using Lorentz-Heaviside units, $\epsilon_0=\mu_0=1$
the parts of the Lagrangian depending on the ALP and photon fields can be written as
\begin{equation}\label{eq:L_a}
  {\cal L}_{a\gamma}=-\frac{1}{4}F_{\mu\nu}F^{\mu\nu}+\frac{1}{2}\partial_\mu a\partial^\mu a+
  \frac{\alpha_{\rm em}}{8\pi}\frac{C_{a\gamma}}{f_a}\,a\,F_{\mu\nu}\tilde F^{\mu\nu}-V_a(a)\,,
\end{equation}
where $F_{\mu\nu}$ is the electromagnetic field strength tensor, $\tilde F_{\mu\nu}$ is its dual,
$C_{a\gamma}$ is a model dependent dimensionless number, and $V_a(a)$ is the
effective ALP potential which can be expanded as $V_a(a)=\frac{1}{2}m_a^2a^2+{\cal O}(a^3)$ around $a=0$.

ALPs can then contribute to cold dark matter as the phase $\theta$ freezes out at a random value of order one.
Other contributions can result from cosmic strings that are formed when the $U(1)$ symmetry associated with the field $a$
is broken. For a review on computing ALP relic densities see Ref.~\cite{Arias:2012az}. Obtaining the correct order of magnitude for the relic
dark matter density requires $m_a\gtrsim10^{-6}\,$eV in the case of axions,

The ALP-photon coupling in Eq.~(\ref{eq:L_a}) couples two photons (which can also be off-shell) to one ALP.
This provides many possible experimental and observational tests for the existence of ALPs.
In {\it light shining through walls experiments} a laser beam is partly converted to ALPs in a strong magnetic field
in front of a wall which is then reconverted by a similar magnetic field within a high Q−value optical cavity.
For example, the Axion-Like Particle Search, alternatively called Any Light Particle Search (ALPS) [949] is operated at DESY
and uses a 5T magnetic field and an optical cavity of 8.4m length~\cite{Ehret:2010mh,ALPS-II}, and the OSQAR experiment at CERN
has recently started~\cite{Ballou:2015cka}.
Photons within stars can be converted to ALPs in the ambient magnetic fields. On the one hand, this leads to
an additional energy loss mechanism that has been used to constrain $f_a$ and $m_a$, see Ref.~\cite{Raffelt:2006cw} for
a recent review, where for axions one obtains $m_a\lesssim10^{-3}\,$eV.
In fact, there are recent hints for extra cooling in certain stellar objects, see Ref.~\cite{Giannotti:2017hny} for a review.
On the other hand, electronvolt scale ALPs emitted from the Sun in this way can be reconverted to X-ray photons in
a strong magnetic field in a dark cavity which is used in so-called {\it helioscopes} such as CAST~\cite{Arik:2011rx,Anastassopoulos:2017ftl}
and its planned successor IAXO~\cite{Giannotti:2016drd}. Further astrophysical tests include core collapse supernova explosions~\cite{Payez:2014xsa},
cosmic microwave background distortions~\cite{Schlederer:2015jwa}, a possible anomalous
transparency of the Universe to $\gamma-$rays~\cite{Meyer:2013pny} and spectral distortions of astrophysical sources~\cite{Conlon:2017qcw,Conlon:2017ofb}.

Finally, if ALPs contribute significantly to the cold dark matter, they can be converted to photons in a strong magnetic field within a dark
cavity. These are called {\it haloscopes} examples of which are ADMX~\cite{Asztalos:2009yp} which scans the mass range
between 1.9$\mu$eV and 3.7$\mu$eV,
and planned future experiments such as MADMAX~\cite{Majorovits:2016yvk} and BRASS which use layered dielectrica in a strong magnetic field.
For $10^{-6}\,{\rm eV}\lesssim m_a\lesssim10\,$eV current constraints can be roughly summarised by $g_{a\gamma}\lesssim10^{-10}\,{\rm GeV}^{-1}$.

The haloscope effect can also occur in astrophysical magnetic fields where it can lead to radio emission from strongly
magnetized astrophysical objects. For the case of resonant conversions in essentially homogeneous magnetic fields
around neutron stars this has been considered in Ref.~\cite{Pshirkov:2007st}, and for non-resonant transitions from around the
Galactic center in Ref.~\cite{Kelley:2017vaa}. In the present paper we estimate the radio fluxes from non-resonant conversions and discuss
their prospects for detection more systematically.

In the next section we derive general expressions for diffuse fluxes and fluxes from discrete sources. In section~\ref{sec:application}
we apply these expressions to concrete astrophysical cases and in section~\ref{sec:detectability} we compare the predicted
fluxes with the sensitivities of present and future radio telescopes. We conclude in section~\ref{sec:Discussion}.

\section{Conversion of axion-like particles into photons in ambient magnetic fields} \label{sec:conversion}
The ALP-photon coupling term in Eq.~(\ref{eq:L_a}) can also be written as
\begin{equation}\label{eq:L_a2}
  \frac{\alpha_{\rm em}}{8\pi}\frac{C_{a\gamma}}{f_a}\,a\,F_{\mu\nu}\tilde F^{\mu\nu}=
  \frac{e^2}{32\pi^2}\frac{C_{a\gamma}}{f_a}\,a\,F_{\mu\nu}\tilde F^{\mu\nu}
  =\frac{\alpha_{\rm em}}{8\pi}\frac{C_{a\gamma}}{f_a}\,a\,F_{\mu\nu}\tilde F^{\mu\nu}
  =\frac{g_{a\gamma}}{4}\,a\,F_{\mu\nu}\tilde F^{\mu\nu}\,,
\end{equation}
where $\alpha_{\rm em}=e^2/(4\pi\epsilon_0)$ and
\begin{equation}\label{eq:g_a}
  g_{a\gamma}\equiv\frac{\alpha_{\rm em}C_{a\gamma}}{2\pi f_a}\,.
\end{equation}
Note that whereas $e^2F_{\mu\nu}\tilde F^{\mu\nu}$ is independent of the electromagnetic units,
$\alpha_{\rm em}F_{\mu\nu}\tilde F^{\mu\nu}$ and thus $g_{a\gamma}$ is not. We will furthermore generally
use units in which $k_{\rm B}=c_0=\hbar=1$.

We now consider the Primakoff effect, the conversion of an ALP of energy-momentum $(E_a,{\bf k}_a)$ and mass $m_a$ into a photon
of energy-momentum $(\omega_\gamma,{\bf k}_\gamma)$ in an external magnetic field whose
Fourier transform is defined by
\begin{equation}\label{eq:Bwk}
  {\bf B}(\omega,{\bf k})=\frac{1}{(2\pi)^2}\int dtd^3{\bf r}\,{\bf B}(t,{\bf r})e^{i(\omega t-{\bf k}\cdot{\bf r})}\,.
\end{equation}
Energy-momentum conservation requires
\begin{equation}\label{eq:conservation}
  E_a=(m_a^2+{\bf k}_a^2)^{1/2}=\omega_\gamma-\omega=(\omega_{\rm pl}^2+{\bf k}_\gamma^2)^{1/2}-\omega
  \,,\quad {\bf k}_a={\bf k}_\gamma-{\bf k}\,,
\end{equation}
where for an electron density $n_e$ the plasma frequency is given by
\begin{equation}\label{eq:omega_pl}
  \omega_{\rm pl}=\left(\frac{e^2n_e}{\epsilon_0m_e}\right)^{1/2}\simeq1.3\times10^3\left(\frac{n_e}{{\rm
        cm}^{-3}}\right)^{1/2}\,{\rm rad}\,{\rm s}^{-1}\,.
\end{equation}
One can then show that for non-relativistic ALPs, $|{\bf k}_a|\ll m_a, k_\gamma$, the conversion rate can be
written as
\begin{eqnarray}
  R_{a\to\gamma}&=&\frac{\pi\epsilon_0}{2}g^2_{a\gamma}n_a\int\frac{d\omega}{T}d^3{\bf k}_\gamma\delta(\omega+E_a-\omega_\gamma)
  \sum_\lambda\left|{\bf B}(\omega,{\bf k}_\gamma-{\bf k}_a)\cdot\bepsilon_\lambda({\bf k}_\gamma)\right|^2\nonumber\\
  &=&\frac{\pi\epsilon_0}{2}g^2_{a\gamma}n_a\frac{1}{T}\int d^3{\bf k}_\gamma
  \sum_\lambda\left|{\bf B}(\omega_\gamma-E_a,{\bf k}_\gamma-{\bf k}_a)\cdot\bepsilon_\lambda({\bf k}_\gamma)\right|^2\,,\label{eq:Rag}
\end{eqnarray}
where $n_a$ is the ALP number density, assumed to be spatially homogeneous, the sum over $\lambda$ is over the two photon polarization states represented
by the vector $\bepsilon_\lambda({\bf k})$, and $T$ is the time scale over which the integration in Eq.~(\ref{eq:Bwk}) is performed.
Note that due to the Wiener-Chintschin theorem $|{\bf B}(\omega,{\bf k})|\propto T^{1/2}V^{1/2}$ with $V$ the volume over which
is integrated in Eq.~(\ref{eq:Bwk}). Therefore, $R_{a\to\gamma}$ is independent of $T$ and proportional to $V$, as it should. Similar
expressions have been discussed in Ref.~\cite{Kelley:2017vaa}. Note that for $\omega_{\rm pl}=m_a$ there is a contribution
from $\omega_\gamma=E_a$, ${\bf k}_\gamma={\bf k}_a$ which leads to a resonance in as static and homogeneous magnetic field.
In the following we will, however, assume dilute plasmas in which the plasma mass can be neglected,
$\omega_\gamma=|{\bf k}_\gamma|=k_\gamma=2\pi\nu$. According to Eq.~(\ref{eq:omega_pl}), when considering photons with
$\omega_\gamma\gtrsim10\,$MHz this is justified for $n_e\lesssim2.3\times10^9\,(\nu/10\,{\rm MHz})^2\,{\rm cm}^{-3}$.

If the magnetic field contains magnetohydrodynamic waves with dispersion relations $\omega\simeq v_mk$ with $v_m$ a characteristic
velocity scale, Eq.~(\ref{eq:Rag}) shows that the photon energies are concentrated around $k_\gamma\simeq m_a$ with a characteristic
relative width
\begin{equation}\label{eq:width}
\Delta\equiv\frac{\Delta k_\gamma}{k_\gamma}\simeq v_a^2/2+v_m+\Delta v\sim10^{-3}\,,
\end{equation}
where $v_a=k_a/m_a\sim10^{-3}$ is the characteristic ALP velocity in the Galaxy and $\Delta v$ is the velocity dispersion within
the object considered which is $\Delta v\sim10^{-3}$ for the Galactic objects we will consider. In a turbulent magnetized medium $v_m$
is of the order of the Alfv\'en velocity which is itself of the order of $\Delta v$.
The last term in Eq.~(\ref{eq:width}) results from the Doppler effect due to relative motion between the dark matter, magnetic field and observer.
Since $v_m\lesssim\Delta v$ the observational signature is thus a line-like photon spectrum with relative width $\Delta\simeq10^{-3}$.

For order of magnitude estimates we will usually consider the static limit in which
$$
  |{\bf B}(\omega,{\bf k})|^2\to\frac{1}{2\pi}2\pi T\delta(\omega)|{\bf B}({\bf k})|^2\,,
$$
with the static Fourier transform defined by
\begin{equation}\label{eq:Bk}
  {\bf B}({\bf k})=\frac{1}{(2\pi)^{3/2}}\int d^3{\bf r}{\bf B}({\bf r})e^{-i{\bf k}\cdot{\bf r}}\,.
\end{equation}
In this limit Eq.~(\ref{eq:Rag}) turns into
\begin{equation}\label{eq:Rag_stat}
  R_{a\to\gamma}=\frac{\pi\epsilon_0}{2}g^2_{a\gamma}n_a\int d^3{\bf k}_\gamma\delta(k_\gamma-E_a)
  \sum_\lambda\left|{\bf B}({\bf k}_\gamma-{\bf k}_a)\cdot\bepsilon_\lambda({\bf k}_\gamma)\right|^2\,.
\end{equation}
This result has been first derived in Ref.~\cite{Sikivie:1983ip} in a slightly different notation.
Let us now approximate $\sum_\lambda\left|{\bf B}({\bf k}_\gamma-{\bf k}_a)\cdot\bepsilon_\lambda({\bf k}_\gamma)\right|^2\simeq B^2({\bf k}_\gamma-{\bf k}_a)$
and express the latter in terms of the magnetic field power spectrum. Assuming homogeneity and isotropy one can write the
magnetic field energy density as
\begin{equation}\label{eq:Mk}
  \rho_m=\frac{1}{2\mu_0V}\int d^3{\bf r}|{\bf B}({\bf r})|^2=\frac{1}{2\mu_0V}\int d^3{\bf k}|{\bf B}({\bf k})|^2=\int d\ln k\rho_m(k)\,,
\end{equation}
thus
\begin{equation}\label{eq:Mk2}
  \rho_m(k)=\frac{2\pi}{\mu_0V}k^3|{\bf B}({\bf k})|^2\,.
\end{equation}
This allows to rewrite Eq.~(\ref{eq:Rag_stat}) as
\begin{equation}\label{eq:Rag_stat2}
  R_{a\to\gamma}\simeq\pi g^2_{a\gamma}\frac{M_a}{m^2_a}\rho_m(m_a)\,,
\end{equation}
where we have taken the non-relativistic approximation $|{\bf k}_\gamma-{\bf k}_a|\simeq k_\gamma\simeq m_a$
and expressed $n_a$ in terms of the total ALP mass within volume $V$, $M_a\simeq n_am_aV$.

We will now apply this formula to the case of diffuse emission and to discrete sources emitting into a small
angular range. For diffuse emission Eq.~(\ref{eq:Rag_stat2}) gives the specific intensity
\begin{equation}\label{eq:Rag_diff}
  I\simeq\pi\frac{g^2_{a\gamma}}{m_a^2}\frac{1}{\Delta}\int_{\rm l.o.s.}dl\rho_a(l)\rho_m(m_a,l)\,,
\end{equation}
where the integral is along the line of sight and we have taken into account that both the ALP mass density $\rho_a=n_am_a$
and the magnetic field power density may depend on the position along the line of sight and $\Delta$ is the
relative photon line width estimated in Eq.~(\ref{eq:width}).

For a discrete source at distance $d$ we get for the total flux density
\begin{equation}\label{eq:Rag_disc}
  S\simeq\frac{\pi}{4d^2}\frac{g^2_{a\gamma}}{m_a^2}\frac{1}{\Delta}\int d^3{\bf r}\rho_a({\bf r})\rho_m(m_a,{\bf r})
  \simeq\frac{\pi}{4d^2}\frac{g^2_{a\gamma}}{m_a^2}\frac{1}{\Delta}M_a\rho_m(m_a)\,,
\end{equation}
where in the last step we have assumed the ALP and magnetic field densities to be roughly constant with $M_a$
the total ALP mass within the object. If the discrete source covers a solid angle $\Omega_s\simeq\pi(r_s/d)^2$ with
$r_s$ the radius of the source, one can also express the flux density as a specific intensity $I=S/\Omega_s$,
\begin{equation}\label{eq:Rag_disc2}
  I\simeq\frac{1}{4r_s^2}\frac{g^2_{a\gamma}}{m_a^2}\frac{1}{\Delta}M_a\rho_m(m_a)\,,
\end{equation}
which does not depend on the distance to the source.

\section{Application to Astrophysical Sources}\label{sec:application}
The relations between photon frequency, wavenumber and ALP mass are given by
\begin{equation}\label{eq:m_an}
  \nu=\omega_\gamma/(2\pi)=242\,\left(\frac{m_a}{\mu{\rm eV}}\right)\,{\rm MHz}\,,\quad
  \frac{1}{k}=20\,\left(\frac{m_a}{\mu{\rm eV}}\right)^{-1}\,{\rm cm}\,.
\end{equation}
We are interested in the ALP mass range $10^{-6}\,{\rm eV}\lesssim m_a\lesssim10^{-3}\,$eV because this corresponds
to the frequency range in which radio telescopes are sensitive and for axions this is the characteristic window in which they
are good candidates for cold dark matter.
For the magnetic field power spectrum we make a power law ansatz
\begin{equation}\label{eq:rho_m}
  \rho_m(k)=\frac{B^2}{2\mu_0}f(k)\,,
\end{equation}
where $B$ is the characteristic total r.m.s. field strength and $f(k)$ is the fraction of the magnetic field energy within
one decade around wavenumber $k$. For turbulent magnetic fields this can be approximated by
\begin{equation}\label{eq:rho_m2}
  f(k)\simeq\left(kl_c\right)^n\,,
\end{equation}
where $l_c$ is the magnetic field coherence length
and $n$ is a spectral index which for Kolmogorov turbulence would be $n=-2/3$. Turbulence is expected to extend
between the scale $l_c$ and the resistive scale which is thought to be fractions of centimeters and thus smaller than
the scale Eq.~(\ref{eq:m_an}). Since in the astrophysical context
$kl_c$ is typically very large, the factor $f(k)$ will be an important limiting factor with a large uncertainty.

Let us also estimate the conversion rate for a single ALP. It is obtained by dividing Eq.~(\ref{eq:Rag_stat2}) by the
total number of ALPs within volume $V$, $N_a=M_a/m_a$ and using the ansatz Eq.~(\ref{eq:rho_m}),
\begin{equation}\label{eq:R_single}
  \frac{1}{\tau_a}\simeq\pi g^2_{a\gamma}\frac{1}{m_a}\rho_m(m_a)\simeq9.7\times10^{-29}\,
  \left(g_{a\gamma}10^{14}\,{\rm GeV}\right)^2\left(\frac{m_a}{\mu{\rm eV}}\right)^{-1}\left(\frac{B}{{\rm G}}\right)^2f(m_a)
  \,{\rm s}^{-1}\,,
\end{equation}
This shows that ALPs will not significantly convert within the age of the Universe unless $g_{a\gamma}$ is close to its
current experimental upper limit, $g_{a\gamma}\lesssim10^{-10}\,{\rm GeV}^{-1}$, and/or the magnetic fields at scale $m_a$
are much stronger than Gauss. It is nevertheless interesting to see what happens if the rate Eq.~(\ref{eq:R_single}) becomes faster
than ALPs can be replaced. In an object of linear size $r_s$ this happens if $r_s/\tau_a\gtrsim v_a$, in numbers
\begin{equation}\label{eq:rep_cond}
  B\gtrsim5.6\times10^{14}\,\left(g_{a\gamma}10^{14}\,{\rm GeV}\right)^{-1}\left(\frac{m_a}{\mu{\rm eV}}\right)^{1/2}
  \left(\frac{10^6\,{\rm cm}}{r_s}\right)^{1/2}\frac{1}{f(m_a)}\,{\rm G}\,.
\end{equation}
In this case the total flux density cannot exceed the limit
\begin{equation}\label{eq:S_max}
  S_{\rm max}\simeq\frac{\rho_a}{m_a}\frac{v_a}{\Delta}\left(\frac{r_s}{d}\right)^2\simeq
  10^{-10}\,\left(\frac{m_a}{\mu{\rm eV}}\right)^{-1}\left(\frac{r_s}{10^6\,{\rm cm}}\right)^2\left(\frac{d}{{\rm kpc}}\right)^{-2}\,{\rm Jy}\,,
\end{equation}
where $d$ is again the distance to the object and $r_s$ is the maximal length scale over which Eq.~(\ref{eq:rep_cond})
applies and the numbers apply as long as the ALP density is not significantly enhanced
in such objects. Eq.~(\ref{eq:S_max}) is in particular relevant for strongly magnetized neutron stars
which have been considered in Ref.~\cite{Pshirkov:2007st} and for which $r_s\sim10\,$km. Flux densities from such
small objects are therefore unlikely to be detectable even with next generation telescopes.

The specific intensity is often expressed in terms of the brightness temperature,
\begin{equation}\label{eq:T_b}
  T_b(\nu)\equiv\frac{c_0^2I}{2\nu^2}\,,
\end{equation}
which in our case for $\nu\simeq m_a/(2\pi)$ gives
\begin{equation}\label{eq:T_b2}
  T_b(m_a)\equiv2\pi^2c_0^2I/m_a^2=0.56\,\left(\frac{I}{{\rm Jy/sr}}\right)\left(\frac{m_a}{\mu{\rm eV}}\right)^{-2}\,{\rm mK}\,.
\end{equation}

Let us now compute numerical estimates for these quantities. For the Galactic diffuse emission
we obtain from Eq.~(\ref{eq:Rag_diff}) for the specific intensity
\begin{eqnarray}
  I&\simeq&1.8\,\left(g_{a\gamma}10^{14}\,{\rm GeV}\right)^2\left(\frac{m_a}{\mu{\rm eV}}\right)^{-2}\left(\frac{10^{-3}}{\Delta}\right)
  \left(\frac{\rho_a}{0.3\,{\rm GeV}{\rm cm}^{-3}}\right)\left(\frac{L}{8\,{\rm kpc}}\right)\nonumber\\
  &&\times\left(\frac{B}{5\,\mu{\rm G}}\right)^2f(m_a)\,\frac{{\rm mJy}}{{\rm sr}}\,,\label{eq:I_gal}
\end{eqnarray}
where $L$ is the characteristic linear size of the Milky Way and for $\rho_a$ we have substituted the canonical local dark
matter density. Inserting this into Eq.~(\ref{eq:T_b2}) gives brightness temperatures in the micro Kelvin range for the fudge factors,
times $f(m_a)$. However, this latter factor is likely to be very small: The typical coherence length of Galactic magnetic
fields is $l_c\sim$pc so that for a Kolmogorov power spectrum $f(m_a)\simeq\left(m_al_c\right)^n\lesssim10^{-13}$.
Even for dark matter and magnetic field profiles that are enhanced toward the Galactic center the total flux is not enhanced by
much more than an order of magnitude. It is thus unlikely that this diffuse emission is detectable in the foreseeable future.

Let us now turn to relatively compact objects which could be small enough for the factor $f(m_a)$
not to be too small. For a discrete source from Eq.~(\ref{eq:Rag_disc}) we obtain for the total flux density
\begin{equation}
  S\simeq2.8\times10^{-11}\,\left(g_{a\gamma}10^{14}\,{\rm GeV}\right)^2\left(\frac{m_a}{\mu{\rm eV}}\right)^{-2}\left(\frac{10^{-3}}{\Delta}\right)
  \left(\frac{M_a}{10^{-10}\,M_\odot}\right)\left(\frac{d}{{\rm kpc}}\right)^{-2}
  \left(\frac{B}{{\rm G}}\right)^2f(m_a)\,{\rm Jy}\,.\label{eq:S_compact}
\end{equation}
Inserting the corresponding specific intensity $I=S/\Omega_s$ into Eq.~(\ref{eq:T_b2}) then gives
for the distance independent brightness temperature
\begin{equation}
  T_b\simeq5\,\left(g_{a\gamma}10^{14}\,{\rm GeV}\right)^2\left(\frac{m_a}{\mu{\rm eV}}\right)^{-4}\left(\frac{10^{-3}}{\Delta}\right)
  \left(\frac{M_a}{10^{-10}\,M_\odot}\right)\left(\frac{r_s}{{\rm pc}}\right)^{-2}
  \left(\frac{B}{{\rm G}}\right)^2f(m_a)\,{\rm nK}\,.\label{eq:Tb_compact}
\end{equation}

The Crab nebula at a distance $d\simeq2\,$kpc has a radius $r_s\simeq2\,$pc and a field $B\sim10^{-3}\,$G. This implies
$M_a\simeq0.3\,M_\odot$. Inserting this into the above formulae yields
\begin{eqnarray}
  S&\simeq&2.1\times10^{-8}\,\left(g_{a\gamma}10^{14}\,{\rm GeV}\right)^2\left(\frac{m_a}{\mu{\rm eV}}\right)^{-2}\left(\frac{10^{-3}}{\Delta}\right)
  \left(\frac{d}{2\,{\rm kpc}}\right)^{-2}\left(\frac{B}{10^{-3}\,{\rm G}}\right)^2f(m_a)\,{\rm Jy}\,,\nonumber\\
  T_b&\simeq&3.8\,\left(g_{a\gamma}10^{14}\,{\rm GeV}\right)^2\left(\frac{m_a}{\mu{\rm eV}}\right)^{-4}\left(\frac{10^{-3}}{\Delta}\right)
  \left(\frac{B}{10^{-3}\,{\rm G}}\right)^2f(m_a)\,\mu{\rm K}\,.\label{eq:Crab}
\end{eqnarray}
Unfortunately, the Crab nebula is very bright at radio frequencies, of the order of $10^3\,$Jy, corresponding to a brightness temperature
of $\simeq10^5\,$K. This would make it very difficult
to extract this small line signal from this large astrophysical foreground, unless $g_{a\gamma}\gtrsim10^{-9}\,{\rm GeV}^{-1}$
even if $f(m_a)$ is not much smaller than one. On the other hand, Wolf-Rayet stars can produce stellar winds
with parameters similar to the Crab nebula, and are also thought to accelerate high energy cosmic
rays~\cite{voelk_biermann,Biermann:1993wg}. In more detail, in an expanding and rotating stellar wind conservation of
angular momentum leads to magnetic breaking of the plasma which in turn can cause the coherent magnetic field $B_c$ to obtain the topology of a
Parker spiral for which $B_c(r)r\simeq$ const. If the turbulent magnetic field component is comparable in strength, then $B(r)r\simeq B(r_s)r_s=$ const.
with $r$ the distance from the source center, then
$\int_0^{r_s}drr^2\rho_aB^2(r)\simeq\rho_aB^2(r_s)r_s^3\simeq M_a(r_s)B^2(r_s)\propto r_s$ for a constant ALP density so that
$M_a(r_s)=4\pi\rho_a r_s^3/3$. Here, $r_s$ can be
identified with the radius of the wind termination shock and can be estimated by equating the mass swept up from the interstellar medium
with the ejected mass $M_e$ which gives
\begin{equation}\label{eq:R_ejecta}
  r_s\sim\left(\frac{3M_e}{4\pi m_Nn_0}\right)^{1/3}\simeq2.1\,\left(\frac{M_e}{M_\odot}\right)^{1/3}
  \left(\frac{1\,{\rm cm}^{-3}}{n_0}\right)^{1/3}\,{\rm pc}\,,
\end{equation}
where $n_0$ is the baryon number density of the interstellar medium and $m_N$ the nucleon mass.
With $B(r)r\lesssim10^{16}\,$Gcm this gives $B(r_s)\lesssim2\times10^{-3}\,$G, and thus similar to the Crab nebula case. Another way
to estimate the magnetic field is by assuming rough equipartition between the kinetic wind and the magnetic field energies which gives
\begin{equation}\label{eq:B_est}
  B(r_s)\sim\left(\frac{3\mu_0M_e}{4\pi r_s^3}\right)^{1/2}v_w\simeq\left(\mu_0m_Nn_0\right)^{1/2}v_w\simeq1.9\times10^{-3}\,
  \left(\frac{n_0}{1\,{\rm cm}^{-3}}\right)^{1/2}\left(\frac{v_w}{10^{-2}}\right)\,{\rm G}\,,
\end{equation}
where $v_w$ is the wind velocity and we have used Eq.~(\ref{eq:R_ejecta}) in the second step. This gives thus values very similar
to the numbers above. Stellar winds from
Wolf-Rayet stars have been observed with flux densities below 0.1 Jy at $d\simeq1\,$kpc, corresponding to brightness temperatures of order a few degrees,
and thus have much lower electron acceleration efficiency than Crab type supernova remnants~\cite{Biermann:1993wg}.
This would considerably simplify the search for a line signal.

We also note that galaxy clusters predict numbers similar to Eq.~(\ref{eq:Crab}) but since coherence scales are likely larger
than parsecs, the suppression factor $f(m_a)$ is likely to be dramatic again.

Let us now try to estimate the suppression factor $f(m_a)$ for such compact objects in a bit more detail.
The magnetic field coherence length in such objects is likely much smaller than for the Galactic magnetic fields.
The Weibel instability may produce magnetic fields on length scales of the Debye length. If one component
of the medium is given by accelerated relativistic electrons and the other by free electrons of density $n_e$
at temperature $T_e$ it is given by
\begin{equation}\label{eq:Debye}
  \lambda_D=\frac{\bar v}{\omega_{\rm
      pl}}\simeq\left(\frac{\epsilon_0T_e}{e^2n_e}\right)^{1/2}\simeq6.9\times10^3
   \left(\frac{T_e}{10^6\rm K}\right)^{1/2}\,\left(\frac{{\rm
      cm}^{-3}}{n_e}\right)^{1/2}\,{\rm cm}\,,
\end{equation}
with $\bar v$ the thermal velocity.
The Bell instability can amplify magnetic fields on the scale of the gyro radius of
cosmic rays of momentum $p$ and charge $Z$ which is given by
\begin{equation}\label{eq:Bell}
  r_g\sim3\times10^{9}\,\left(\frac{10^{-3}{\rm G}}{B}\right)\,\left(\frac{p/Z}{{\rm GeV}}\right)\,{\rm cm}\,.
\end{equation}
In any case, if cosmic rays are accelerated in such objects significant magnetic power on such
scales is required to enable diffusive acceleration. The scales in Eqs.~(\ref{eq:Debye}) and~(\ref{eq:Bell}) are
larger than the scale Eq.~(\ref{eq:m_an}) only by a few orders of magnitude so that the suppression factor $f(m_a)$
may be moderate.

We also note that for $\nu\gtrsim1\,$MHz free-free absorption is generally negligible within the Galaxy, as can
be seen from the absorption rate which in the Rayleigh-Jeans regime $\nu\la k_{\rm B}T_e/h=2.1\times10^{13}(T_e/10^3\,{\rm K})\,$Hz
is given by
\begin{equation}
  \alpha^{\rm ff}_\nu\simeq9\times10^{-4}\left(\frac{n_e}{0.1\,{\rm cm}^{-3}}\right)^2
  \left(\frac{10^3\,{\rm K}}{T_e}\right)^{3/2}\left(\frac{{\rm MHz}}{\nu}\right)^2\,{\rm pc}^{-1}\,,\label{eq:a_ff3}
\end{equation}
where $T_e\gtrsim10^3\,$K is the temperature of the medium.

\section{Detectability}\label{sec:detectability}
The effective solid angle of a single Gaussian beam is given by
\begin{equation}\label{eq:Omega_b}
  \Omega_b\simeq\theta^2\simeq\frac{1}{(l\nu)^2}=\frac{1}{A\nu^2}\,,
\end{equation}
where $\theta$ is the angular radius of the beam, $l$ is the effective length scale of the interferometer
and $A=l^2$ its effective area. If a discrete source extends over several beams, the sensitivity in brightness
temperature is increased by a factor $N^{1/2}_b=(\Omega_s/\Omega_b)^{1/2}$ relative to a
single beam so that the minimal detectable brightness temperature is given by
\begin{equation}\label{eq:T_b_sens}
  T_{b,{\rm min}}\simeq \frac{T_{b,{\rm min0}}}{N^{1/2}_b}=T_{b,{\rm min0}}\left(\frac{\Omega_b}{\Omega_s}\right)^{1/2}\,,
\end{equation}
where $T_{b,{\rm min0}}$ is the sensitivity for a single beam. In general one has
\begin{equation}\label{eq:T_b_sens2}
  T_{b,{\rm min0}}\simeq\frac{T_{\rm noise}}{(Bt)^{1/2}}\,,
\end{equation}
where $T_{\rm noise}$ is the effective noise temperature, resulting from system and sky temperature added in quadrature,
$B$ is the bandwidth and $t$ is the observing time.
One also often uses the antenna temperature induced by a total flux density $S$ defined by
\begin{equation}\label{eq:T_a}
  T_a\equiv\frac{AS}{2}=0.36\,\left(\frac{A}{10^3\,{\rm m}^2}\right)\left(\frac{S}{{\rm Jy}}\right)\,{\rm K}\,.
\end{equation}
Combining this with Eqs.~(\ref{eq:T_b}) and~(\ref{eq:Omega_b}) and the relation $S=I\Omega_s$ this shows that
\begin{equation}\label{eq:T_a2}
  \frac{T_a}{T_b}=N_b=\frac{\Omega_s}{\Omega_b}\,.
\end{equation}
If the noise in one beam is again characterized by the temperature $T_{b,{\rm min0}}$, the noise in $N_b$
beams corresponds to $N^{1/2}_bT_{b,{\rm min0}}$. Comparing this with the total signal temperature $T_a$
again gives a brightness temperature sensitivity improvement by a factor $N^{1/2}_b$. Equivalently, the smallest
detectable total source flux density can be expressed as
\begin{equation}\label{eq:S_min}
  S_{\rm min}=N^{1/2}_bS_b\,,
\end{equation}
where $S_b$ is the minimal detectable flux density per beam. Since $S_{\rm min}$ and $S_b$ are proportional to
$T_{b,{\rm min0}}$ which according to Eq.~(\ref{eq:T_b_sens2}) is proportional to $1/t^{1/2}$, one often denotes the minimal detectable
source flux density in units of Jy$\,{\rm hr}^{-1/2}$.

Let us now apply these estimates to various relevant experiments. LOFAR HBA~\cite{Shimwell:2016xsq} has a beam size of $\simeq5\,$arcsec
which thus covers a solid angle $\Omega_b\simeq2\times10^{-9}\,$sr, at $\nu=140\,$MHz, corresponding to $m_a=0.58\,\mu$eV, see Eq.~(\ref{eq:m_an}).
For the stellar nebulae of radial extent $r_s\simeq2\,$pc at a distance $d$ discussed above this
could increase the sensitivity by a factor $N^{1/2}_b\simeq41\,(2\,{\rm kpc}/d)$. For a sensitivity of $S_b\sim10^{-4}\,$Jy per beam
Eqs.~(\ref{eq:T_b2}) and~(\ref{eq:T_b_sens}) then predict a sensitivity of $T_b\simeq2[d/(2\,{\rm kpc})]\,$K.
Expressed in terms of total source flux density this corresponds to $S_{\rm min}\simeq4\times10^{-3}\,(2\,{\rm kpc}/d)\,$Jy
according to Eq.~(\ref{eq:S_min}). Comparing this with the prediction Eq.~(\ref{eq:Crab}) suggests that couplings
$g_{a\gamma}\gtrsim2.5\times10^{-12}\,[m_a/(0.58\,\mu{\rm eV})][d/(2\,{\rm kpc})]^{1/2}\,{\rm GeV}^{-1}/f(m_a)$ may be testable for $m_a\simeq\,\mu$eV.

The planned SKA-low is sensitive in the frequency range between 50 and 350 MHz. It has a beam size of order of a square degree,
$\Omega_b\sim3\times10^{-4}\,$sr, which is typically larger than the angular size of the
sources we have discussed here so that $N_b=1$. The source flux density sensitivities are of order $10\,\mu{\rm Jy}\,{\rm hr}^{-1/2}$.
According to Eq.~(\ref{eq:T_b2}) in terms of brightness temperature this corresponds to $\simeq10\,\mu{\rm K}\,{\rm hr}^{-1/2}$, see, e.g., Ref.~\cite{ska}.
This can also be seen from Eq.~(\ref{eq:T_b_sens2}) for $T_{\rm noise}\simeq10\,$K and $B\simeq300\,$MHz.
Comparing this with the prediction Eq.~(\ref{eq:Crab}) for $S$ this translates to possible sensitivities down to
$g_{a\gamma}\gtrsim2\times10^{-13}\,[m_a/\mu{\rm eV}][d/(2\,{\rm kpc})]^{1/2}\,{\rm GeV}^{-1}/f(m_a)$ within about an hour of observing time.
The planned SKA-mid is sensitive in the frequency range between 0.35 and 14 GHz and the sensitivity in terms of flux densities is about
a factor 5 lower. On the other hand the predicted $S\propto(g_{a\gamma}/m_a)^2$ so that constraints on $g_{a\gamma}$ degrade by factors of
a few. Note that according to Eqs.~(\ref{eq:m_a}) and~(\ref{eq:g_a}) for the QCD axion one has
$g_{a\gamma}\simeq5\times10^{-15}(m_a/\mu{\rm eV})\,{\rm GeV}^{-1}$.

Up to now we have quoted sensitivities for continuum emission. However, the predicted emission is line-like with a relative
width of $\Delta\sim10^{-3}$, see Eq.~(\ref{eq:width}). This implies that the optimal effective bandwidth has to be set to
\begin{equation}\label{eq:B_width}
  B\simeq\Delta\nu=242\,\left(\frac{m_a}{\mu{\rm eV}}\right)\left(\frac{\Delta}{10^{-3}}\right)\,{\rm kHz}\,.
\end{equation}
This leads to sensitivities that are about a factor $\simeq30(10^{-3}/\Delta)^{1/2}$ worse than the numbers quoted above and thus to limits on
$g_{a\gamma}$ that are degraded by a factor $\simeq6(10^{-3}/\Delta)^{1/4}$, to
$g_{a\gamma}\gtrsim10^{-12}\,[m_a/\mu{\rm eV}][d/(2\,{\rm kpc})]^{1/2}\,{\rm GeV}^{-1}/f(m_a)$
for SKA. On the other hand, since the signal at a given frequency or ALP mass
scales as $1/\Delta$, the signal to noise ratio scales as $1/\Delta^{1/2}$. Overall the sensitivity to $g_{a\gamma}$ scales as $1/\Delta^{1/4}$
and increases with observing time as $t^{1/4}$. Longer observing times and optimised broadband searches may thus increase the sensitivity.

\section{Outlook and Conclusions} \label{sec:Discussion}
Under optimistic assumptions for the magnetic field power spectrum at meter scales we have shown that
strongly magnetized stellar winds may probe ALP-photon coupling parameters below current upper limits
$g_{a\gamma}\lesssim10^{-10}\,{\rm GeV}^{-1}$ in the ALP mass range around $10^{-6}\,$eV through radio
emissions that may be detectable with existing radio telescopes such as LOFAR HBA and with future experiments such as SKA.
These experiments together cover frequencies between $\simeq10\,$MHz and $\simeq15\,$GHz, corresponding to ALP masses
$0.1\,\mu{\rm eV}\lesssim m_a\lesssim100\,\mu$eV.
Around $m_a\sim\mu$eV observations of such discrete astrophysical objects should be sensitive to ALP-photon couplings
$g_{a\gamma}\gtrsim10^{-12}\,[m_a/\mu{\rm eV}][d/(2\,{\rm kpc})]^{1/2}\,{\rm GeV}^{-1}/f(m_a)$
where $f(m_a)<1$ describes the fraction of the magnetic field power on scales $k\simeq m_a$ which for
turbulent spectra can be approximated by $f(m_a)\simeq\left(m_al_c\right)^n$ with $n<0$
and $l_c$ the coherence scale. The strongest constraints thus tend to be given by the most nearby sources.
Furthermore, longer observation times and dedicated analysis methods may increase the sensitivity so that
sensitivities down to $g_{a\gamma}\simeq10^{-13}\,{\rm GeV}^{-1}$ may be reachable.
If laboratory haloscopes would find indications for the existence of ALP dark matter and fix its mass
$m_a$ and coupling scale $g_{a\gamma}$ the observation of a photon line in an astrophysical haloscope would
in turn allow to derive characteristics of the magnetic field in the corresponding astrophysical object, in
particular the power $\rho_m(m_a)$ around wavenumber $m_a$, or the combination $B^2f(m_a)$.

\begin{acknowledgments}
This work has been supported by the Deutsche Forschungsgemeinschaft through the Collaborative Research Center SFB 676 ``Particles,
Strings and the Early Univers'', and by the Helmholtz Alliance for Astroparticle Physics (HAP) funded by the Initiative and
Networking Fund of the Helmholtz Association.
We are grateful to Marcus Br\"uggen for useful comments on the manuscript.

\end{acknowledgments}


\end{document}